\documentclass[letterpaper]{article} % DO NOT CHANGE THIS

\usepackage[preprint]{aaai2027} % DO NOT CHANGE THIS

\usepackage[hyphens]{url} % DO NOT CHANGE THIS
\usepackage{graphicx} % DO NOT CHANGE THIS
\usepackage{natbib} % DO NOT CHANGE THIS
\usepackage{caption} % DO NOT CHANGE THIS
\usepackage{amsmath}
\usepackage{amssymb}

\usepackage{algorithm}
\usepackage{algorithmic}

\usepackage{booktabs}
\usepackage{array}
\usepackage{tabularx}
\usepackage{multirow}
\usepackage{threeparttable}

\usepackage{subcaption}

\usepackage{newfloat}
\usepackage{listings}

\DeclareCaptionStyle{ruled}{
  labelfont=normalfont,
  labelsep=colon,
  strut=off
}

\floatstyle{ruled}
\newfloat{listing}{tb}{lst}{}
\floatname{listing}{Listing}

\newcommand{\system}{ACLE-MCP}
\newcommand{\trustgap}{post-authorization execution trust gap}
\newcommand{\gate}{Execution Gate}
\newcommand{\AR}{\mathsf{AR}}
\newcommand{\CL}{\mathsf{CL}}
\newcommand{\IB}{\mathsf{I}}

\ifdefined
\fi

\title{ACLE-MCP: Attested Capability Leases for Execution-Time Trust in Remote LLM Tool Use}

\author{
Zhiyang Ding,
Yang Luo\textsuperscript{*},
Guangpu Chen,
Qingni Shen,
Zhonghai Wu
}

\affiliations{
Peking University\\
dingzhiyang26@stu.pku.edu.cn,
luoyang@pku.edu.cn,
chenguangpu@stu.pku.edu.cn,\\
qingnishen@ss.pku.edu.cn,
wuzh@pku.edu.cn\\
\textsuperscript{*}Corresponding author
}

\begin{document}

\maketitle

\begin{abstract}
Remote Model Context Protocol (MCP) services enable large language model agents to invoke external tools, but OAuth authorization alone does not ensure that a later tool call is executed by the provider-side workload that the relying party intended to trust. An endpoint may remain authorized even after execution shifts to a substituted workload, relies on stale appraisal state, reuses authority transferred from another sender, or traverses an undeclared downstream component. We call this problem the \emph{post-authorization execution trust gap}. We present \system{}, an invocation-scoped architecture that couples delegated authorization, workload appraisal, and resource-side execution admission. For protected calls, \system{} issues a short-lived, sender-constrained capability lease that binds the expected workload, freshness requirement, operation, object and parameter bounds, downstream constraints, and receipt obligations. A provider-side \gate{} consumes the lease immediately before protected tool logic begins. We implement a runnable prototype with Keycloak/OIDC validation, an MCP Python SDK server, and an optional vTPM quote-verification backend. Controlled security experiments and an agent tool-use extension show that weaker authorization or connect-time attestation modes leave distinct post-authorization attacks open, whereas full \system{} blocks all evaluated attack families while preserving all benign tasks. In the locally simulated agent extension, the complete design increases request-level pooled p95 latency on normal allowed calls by 25.7\% relative to OAuth-only. These results indicate that invocation-time binding between call authority and current workload state is a practical complement to OAuth-protected remote tool use.
\end{abstract}

\section{Introduction}
\label{sec:Introduction}
Large language model (LLM) applications increasingly invoke external tools, data sources, and enterprise services through the Model Context Protocol (MCP). This shift from passive text generation to agentic execution allows a Host to retrieve private data, update records, invoke internal APIs, and trigger external workflows. In remote deployments, OAuth can authorize an MCP Client to access a protected MCP Server. OAuth answers an important question---whether a client or user may access a resource---but it does not establish that the concrete provider-side workload executing a later invocation is still the execution unit that the relying party intended to trust.

We call this problem the \emph{post-authorization execution trust gap}. A remote endpoint may complete OAuth, expose plausible MCP metadata, and return syntactically valid results while the underlying execution unit has been replaced, rolled back, misconfigured, compromised, or routed through an undisclosed downstream component. An attestation result obtained at registration or connection time may also become stale before authority is actually consumed. The authorization decision can therefore remain valid even though the execution context no longer satisfies the trust assumptions under which the call was allowed.

Consider an agent authorized to invoke \texttt{update\_ticket} on ticket \texttt{T-17}. The Host selects the correct tool, validates the arguments, and obtains a valid access token. Before the invocation reaches the protected handler, however, the provider may roll the serving container back to a vulnerable image, route the request through an undeclared proxy, or accept a lease copied from another sender. Operation and parameter bounds remain important, but they are not the central research problem; rather, they ensure that authority issued after successful workload appraisal cannot be widened when consumed. The core question is whether an OAuth-authorized call is still executed by the expected, freshly appraised workload within the bounds for which the additional authority was issued.

Existing mechanisms protect adjacent but distinct security boundaries. Prompt-injection defenses and plan validators operate before remote execution. OAuth and rich authorization mechanisms govern delegated access, but do not identify or appraise the concrete provider-side workload that ultimately consumes authority. Sender-constrained tokens reduce credential-transfer risk, but do not establish current workload state. Remote attestation can establish workload properties, yet an appraisal obtained at connection time may become invalid before a later invocation. What is missing is an invocation-scoped binding among the OAuth authorization context, current workload appraisal, caller sender proof, and resource-side admission decision at the execution boundary.

We present \system{}, an attested capability-lease architecture for high-risk remote MCP invocations. For a protected call, a Verifier appraises evidence from the target workload and returns a freshness-bounded attestation result. A lease issuer combines that result with the verified OAuth context and requested invocation boundary. The resulting short-lived, sender-constrained lease binds workload identity, operation and object scope, parameter bounds, side-effect budget, downstream constraints, freshness, and any receipt obligation. A provider-controlled \gate{} validates and consumes the lease at the final admission point before protected tool logic begins.

We implement a Python prototype with separate Verifier, lease-issuer, gate, and tool services. A realistic integration path validates standard Keycloak/OIDC access tokens and forwards admitted requests to an MCP Python SDK Server. Keycloak remains the ordinary OAuth/OIDC Provider; an independent lease issuer consumes the verified OAuth context and appraisal result to issue \system{} leases. An optional \texttt{swtpm}/\texttt{tpm2-tools} path replaces the simulated evidence signature with nonce-bound quote verification while preserving the same appraisal-result, lease, and gate interfaces.

This paper makes three contributions:
\begin{itemize}
    \item We identify and formalize the \trustgap{} in remote MCP deployments: valid OAuth authorization does not imply that a later tool call is executed by the expected and freshly appraised provider-side workload.

    \item We design \system{}, which binds OAuth-derived authority, sender key possession, workload appraisal, invocation constraints, and provider-side admission into a short-lived capability consumed immediately before execution.

    \item We implement and evaluate five enforcement modes. The evaluation covers replay, workload substitution, stale appraisal, scope misuse, undeclared proxying, and receipt violations, and extends the analysis to chained and streaming agent tool-use scenarios without redefining the core security problem.
\end{itemize}

\section{Security Problem}

\subsection{Post-Authorization Execution Trust Gap}
For an invocation at time $t$, let the Host-authorized invocation boundary be
\begin{equation}
\IB_t=\langle aud,tool,op,obj,\Theta,D,B,\rho\rangle,
\end{equation}
where $aud$ is the intended audience; $tool$, $op$, and $obj$ identify the tool, operation class, and object scope; $\Theta$ specifies parameter constraints; $D$ is the approved downstream set; $B$ is the side-effect budget; and $\rho$ is the minimum workload-assurance policy.

Let the corresponding provider-side execution observation be
\begin{equation}
E_t=\langle wid,aud',tool',op',obj',\theta',D',s,\tau\rangle,
\end{equation}
where $wid$ identifies the workload actually serving the request, $s$ is its appraised state, and $\tau$ is the evidence time. OAuth authorization establishes only that the client may access $aud$; it does not imply that $wid$ is the expected workload, that $s$ satisfies $\rho$, or that the appraisal remains sufficiently fresh at execution time. The \trustgap{} is precisely this missing implication.

ACLE-MCP targets the following admission property:
\begin{equation}
\begin{aligned}
\mathsf{Allow}(E_t)\Rightarrow {}&
\mathsf{OAuthOK} \\
&{}\land \mathsf{FreshWorkload}(wid,s,\tau,\rho) \\
&{}\land \mathsf{WithinBounds}(E_t,\IB_t).
\end{aligned}
\end{equation}
The first two terms express the core contribution: delegated access and current workload trust must hold simultaneously at invocation time. The final term confines authority issued after appraisal so that it cannot be consumed for a broader operation, object, parameter range, side effect, or downstream path. This property neither proves arbitrary tool semantics nor infers latent user intent.

\subsection{Attack Classes and Security Goals}
The core attack classes follow directly from the trust gap. \textbf{Credential replay or transfer} attempts to reuse authority for another sender or invocation. \textbf{Workload substitution} replaces the serving process, container, virtual machine, enclave, image, configuration, or dependency state after authorization. \textbf{Stale appraisal reuse} exploits the interval between an earlier appraisal and a later execution decision. \textbf{Invocation-boundary misuse} consumes a valid lease for a different operation, object, or parameter range. \textbf{Undeclared delegation} forwards execution to a downstream component outside the authorized path. \textbf{Receipt violations} remove or forge the structured execution evidence required for high-impact operations.

The chained, reordered, repeated, and streaming cases in the agent extension are concrete ways in which invocation-scoped authority can be reused or widened. They test the applicability of the lease abstraction in complex agent workflows, but they do not replace the paper's original post-authorization workload-trust problem.

\subsection{Threat, Trust, and Policy Model}
The adversary may control or influence the MCP-facing application workload, its deployment configuration, or an apparently legitimate downstream proxy. The adversary may return syntactically valid responses, replay captured credentials, reuse stale appraisal results, substitute a workload after authorization, broaden request arguments, reuse authority across steps, or forward execution to an undeclared dependency. These capabilities are consistent with documented risks in LLM tool chains, MCP deployments, excessive agency, insecure tool integration, and compromised connected systems~\cite{owasp-llm-top10,nist-genai-profile,cve-2025-52573,cve-2025-6515,mcp-security-survey}.

We trust the Host's invocation-boundary construction logic, the Verifier, the lease issuer, and the provider-side \gate{}. We assume standard cryptographic mechanisms are secure and that the gate is non-bypassable for protected handlers. The application workload is outside the trusted computing base and may be stale, substituted, compromised, or misconfigured. If a Provider exposes an alternate path around the gate, the guarantee does not apply to that path. We do not address prompt injection, incorrect Host-side planning, malicious local \texttt{stdio} launch commands, compromise of the trusted components, or hidden dependencies that are neither declared nor mediated.

Policy has two owners. Host policy determines the user-authorized invocation boundary and cannot be widened by remote tool metadata. Provider policy determines acceptable workload states, appraisal freshness, and deployment-specific downstream constraints. Appraisal results and leases carry signed policy-version identifiers. The lease issuer computes the intersection of the two policies; a conflict, missing version, or policy downgrade fails closed rather than silently reducing assurance.

\section{Related Work}
\label{sec:related-work}
\textbf{Agent and tool security.} ToolHijacker, cross-tool harvesting and pollution, and MCP memory attacks show that malicious tool descriptions, tool outputs, and shared state can corrupt planning and cross-tool behavior~\cite{toolhijacker,xthp,msa}. ACE, SAGA, and AgentSentinel introduce barriers, governance, authentication, or runtime monitoring around agent planning and execution~\cite{ace,saga,agentsentinel}. AgentDojo and InjecAgent provide environments for studying prompt injection against tool-using agents~\cite{agentdojo}. These efforts primarily protect plan formation, tool selection, agent communication, or Host-observable actions. \system{} addresses a later remote boundary: after OAuth authorization succeeds, whether the provider-side workload that consumes the call is still the expected and freshly appraised execution unit.

\textbf{Delegated authorization and capabilities.} OAuth provides delegated access to protected resources. Rich Authorization Requests express structured authorization details, Token Exchange supports delegation, DPoP sender-constrains tokens, and Stateful Least Privilege incorporates Server-side action history into authorization policy~\cite{rar,token-exchange,dpop,stateful-lp}. Macaroons and WAVE demonstrate attenuated, contextual, and delegated capabilities~\cite{macaroons,wave}. \system{} builds on these ideas rather than replacing them. Its additional contribution is to bind narrow invocation authority to a fresh provider-side workload appraisal and consume that authority at the tool-execution boundary.

\textbf{Remote attestation and execution control.} RATS separates Evidence, Appraisal, and Relying Party Decisions~\cite{rats}. Runtime SBOM verification, Attestable Builds, ARI, Full Trust Alchemist, and Transparent Attested DNS strengthen remote workload identity, build provenance, or dynamic appraisal~\cite{runtime-sbom-ra,attestable-builds,ari,full-trust-alchemist,tadns}. Confidential-computing systems such as SHELTER and Acai provide lower-level workload or accelerator isolation~\cite{shelter,acai}. These systems establish or protect execution environments. \system{} instead studies how an appraisal result should change the admission decision for one concrete remote agent tool invocation.

\section{Methodology}
\label{sec:methodology}

\subsection{Design Overview}
\begin{figure*}[t]
  \centering
  \includegraphics[width=0.98\textwidth]{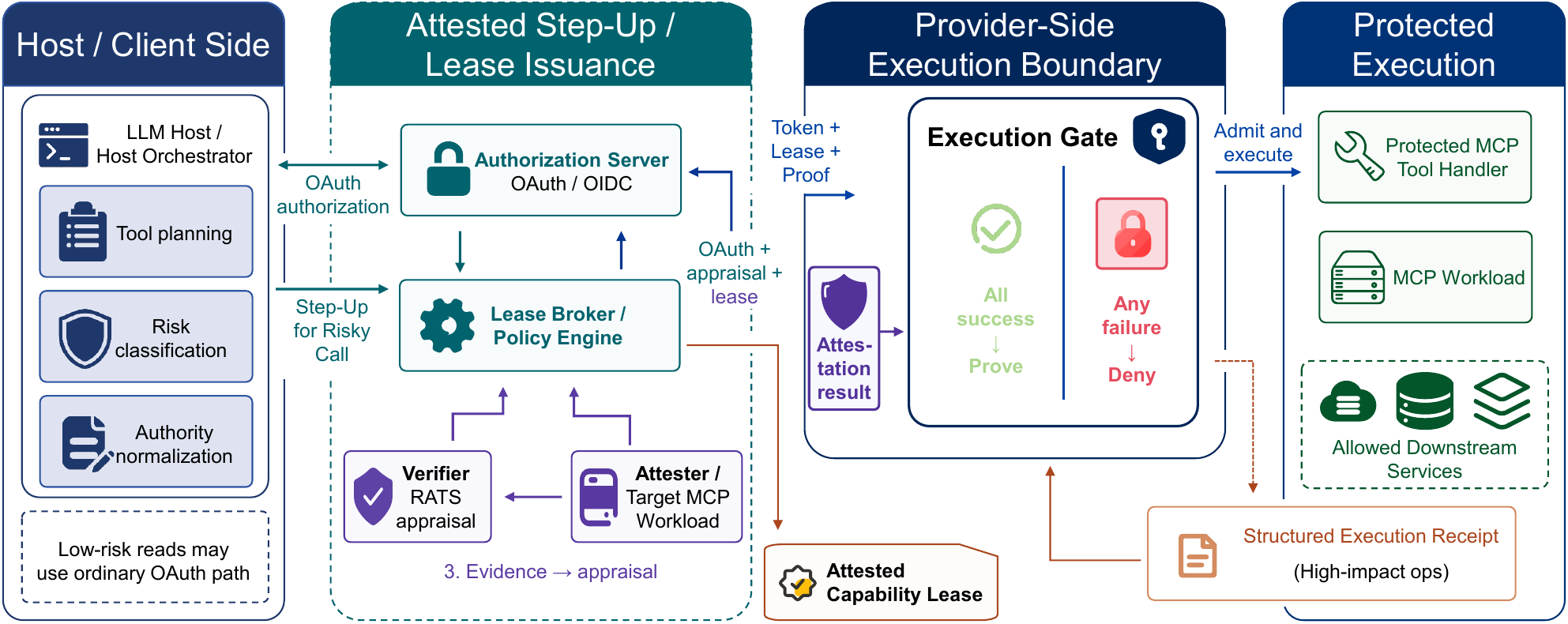}
  \caption{Overview of \system{}. OAuth authorizes the client to access the remote MCP resource; Attested Step-Up binds protected invocation authority to a fresh appraisal of the expected provider-side workload; and the \gate{} validates the relevant credentials and consumes the resulting lease immediately before protected tool logic begins.}
  \label{fig:acle-overview}
\end{figure*}

Figure~\ref{fig:acle-overview} shows how \system{} closes the post-authorization execution trust gap. The Host Orchestrator first normalizes the selected structured call and compiles the invocation boundary $\IB_t$. Low-risk calls may remain on the ordinary OAuth path when both Host and Provider policies permit. Medium- and high-risk calls trigger Attested Step-Up. The Verifier appraises the provider-side workload expected to serve the call, and the lease issuer combines the signed appraisal result, verified OAuth context, sender identity, and $\IB_t$ into a short-lived capability lease. The provider-side \gate{} validates and consumes the lease before protected tool logic begins; only after all checks pass may the application workload execute the call.

The placement of the final check is essential. When OAuth authorization completes, the concrete provider-side execution unit may not yet be known; connect-time appraisal may also precede the protected invocation by an unsafe interval. The lease carries the intersection of the Host-authorized invocation boundary and provider-side workload policy to the point where authority is actually consumed. A request is denied if the sender or serving workload does not match the lease, if the appraisal is stale, or if the tool, operation, object, parameters, side effects, or downstream path exceed the issued boundary.

\subsection{Attested Step-Up and Lease Issuance}
At tool registration, the Host records a local policy profile for each tool, including default operation, data, side-effect, scope, and dependency classes. At invocation time, the boundary compiler refines these defaults using the concrete object, parameter ranges, tenant boundary, external transmission, administrative scope, expected tool-chain position, and declared downstream path. It then produces $\IB_t$ and the minimum workload-assurance requirement $\rho$.

Risk classes directly control enforcement. A low-risk read may continue through ordinary OAuth when Provider policy permits. A medium-risk invocation requires a narrowly scoped lease and an appraisal result within a bounded reuse window. A high-risk invocation requires stricter freshness, a shorter or single-use lease, tighter side-effect bounds, and any execution receipt required by policy. Uncertainty may only raise the risk class; it never triggers an automatic downgrade.

For a protected invocation, an attestation coordination component obtains evidence for the workload expected to serve the call. Following the RATS separation of Evidence, Appraisal, and Relying Party Decisions~\cite{rats}, the Verifier checks evidence integrity, nonce and time freshness, platform- or Provider-specific measurements, and Provider policy. It then emits
\begin{equation}
\AR_t=\mathsf{Sign}_{V}\langle wid,state,ts,exp,pid_P,verdict\rangle.
\end{equation}
Here, $wid$ identifies the appraised workload, $state$ captures policy-relevant state, and $ts$ and $exp$ bound the result's validity interval. Raw evidence may remain inside the Provider--Verifier boundary; the Host, lease issuer, and gate require only the minimized signed result needed for admission.

Keycloak continues to issue standard OAuth/OIDC access tokens. After validating the OAuth context, $\IB_t$, and $\AR_t$, the lease issuer creates
\begin{equation}
\CL_t=\mathsf{Issue}(\IB_t,\AR_t,sender,expiry,pid_H,pid_P).
\end{equation}
The concrete lease binds the issuer, subject, audience, sender confirmation key, workload identity, tool, operation and object, parameter bounds, execution-state constraints, downstream set, side-effect budget, appraisal digest, policy versions, receipt obligation, expiration, and unique lease identifier. OAuth represents delegated access; the lease represents permission to execute one bounded invocation under a particular workload state. The Capability-only baseline retains request caveats but omits fresh workload binding, thereby separating fine-grained authority constraints from invocation-time workload appraisal.

\subsection{Invocation-Time Enforcement}
The \gate{} is placed before the protected handler and controlled by the Provider or platform boundary rather than by the potentially compromised application workload. It may be implemented as a reverse proxy, sidecar, gateway, service-mesh authorization filter, or confidential-computing wrapper, provided alternate paths are prevented. Algorithm~\ref{alg:gate} gives the mandatory enforcement order.

\begin{algorithm}[t]
\caption{Invocation Admission at the Execution Gate}
\label{alg:gate}
\begin{algorithmic}[1]
\REQUIRE Request $q$, access token $AT$, lease $L$, sender proof $P$, appraisal result $AR$
\ENSURE \textsc{Allow} with an optional receipt, or \textsc{Deny}
\STATE Validate $AT$ for issuer, subject, audience, and target resource.
\STATE Validate $L$ for issuer, signature, audience, expiration, and policy versions.
\STATE Verify $P$ against the sender key bound in $L.cnf$.
\STATE Verify $AR$; require $AR.wid=L.wid$ and require $AR$ to be sufficiently fresh for $L$.
\STATE Reject any authority reuse that violates $L$'s replay or execution-state policy.
\STATE Require $q$ to satisfy $L$'s tool, operation, object, parameter, side-effect, and downstream bounds.
\IF{$L$ requires a receipt}
    \STATE Execute only through the receipt-producing protected path.
    \STATE Verify that the receipt binds $L$, $q$, $AR$, the result digest, and the downstream summary.
\ELSE
    \STATE Execute through the protected tool handler.
\ENDIF
\STATE Mark single-use authority as consumed when required and return \textsc{Allow}.
\end{algorithmic}
\end{algorithm}

This order prevents side effects before validation completes. Sender binding prevents a copied lease from becoming portable authority. Workload binding prevents an appraisal for one execution unit from authorizing another. Freshness narrows the attestation-to-execution window. Request and state constraints prevent a lease issued for one tool, object, order, stream state, or downstream set from authorizing another invocation.

Downstream constraints apply only to dependencies visible to enforcement. A Provider may validate declared endpoints, mediate egress traffic, or require a signed dependency manifest. Each protected hop in a chained workflow therefore requires either a separate lease or inclusion in an explicitly authorized downstream set. Receipts for high-impact operations can provide structured accountability, but cannot undo external side effects or prove arbitrary program semantics. A protected call fails closed whenever workload appraisal, lease issuance, sender key possession, freshness, policy-version agreement, boundary validation, or required receipt validation fails.

\section{Evaluation}
\label{sec:Evaluation}
We evaluate three questions: \textbf{RQ1}: Does invocation-time workload binding close the modeled post-authorization trust gap while preserving benign tool calls? \textbf{RQ2}: Which mechanisms account for the observed security gains? \textbf{RQ3}: What integration cost and runtime overhead does the prototype introduce?

\subsection{Experimental Setup}
We implement separate Python services for the Verifier, lease issuer, \gate{}, MCP tool service, and adversarial workloads, with components communicating over HTTP. The main controlled experiments use simulated OAuth, workload appraisal, and sender proof to isolate lease issuance and resource-side execution logic. The Verifier produces signed appraisal results from evidence, freshness, and Provider policy; the lease issuer validates the OAuth context, invocation boundary, and appraisal result before issuing a signed, short-lived, sender-constrained capability lease; and the \gate{} validates the token, lease, sender proof, workload binding, and invocation boundary before the protected handler. The harness records request-level decisions, randomized scenario order, environment manifests, stage-level latency, and security outcomes.

A separate integration path uses Keycloak for OAuth/OIDC authorization and validates access-token issuer, client, and resource audience through the issuer's JWKS. Admitted requests are forwarded to a tool service implemented with the MCP Python SDK. An optional \texttt{swtpm}/\texttt{tpm2-tools} path verifies nonce-bound vTPM quotes over PCRs 0, 1, 2, 3, 4, 7, and 11 to confirm that the Verifier backend is replaceable. This path is not used to measure production TPM performance and does not claim complete hardware binding for every application-level workload property.

We compare five modes. \textbf{OAuth-only} validates bearer access but does not check workload state or invocation bounds. \textbf{Attest-on-connect} records workload appraisal at connection time but does not consume fresh evidence for each invocation. \textbf{Stateful LP} checks tool, operation, object, parameters, execution order, and declared downstream policy at the resource side, but omits sender proof and workload appraisal. \textbf{Capability-only} uses a short-lived capability with audience and request caveats but no fresh workload binding. \textbf{ACLE-MCP} enables the complete enforcement chain.

The agent tool-use extension contains four benign task families: filesystem summarization, GitHub issue commenting, bounded database reporting, and Kubernetes status inspection. It also contains six execution-misuse attack families: wrong-tool execution, validate/execute order reversal, tool-chain step injection, authority reuse across tool-chain steps, downgrade or re-execution, and streaming semantic mutation. Each scenario is repeated ten times under each mode, yielding 500 mode--scenario observations. We report benign-task success, attack-blocking rate, false-positive and false-negative rates, and request-level pooled p95 latency over normal expected-allow requests that were actually allowed. Correctly blocked attack scenarios are security outcomes rather than runtime errors.

\subsection{RQ1: Security Effectiveness and Agent Extension}
The mechanism-level security experiments first validate ACLE-MCP's original claim: OAuth-only and connect-time attestation admit replay, workload substitution, stale appraisal, undeclared proxying, and receipt violations because neither mode binds a protected invocation to the current provider-side appraisal result at the gate. The component results in RQ2 identify the checks required to close these attacks.

Table~\ref{tab:agent-main} reports the agent tool-use extension. All five modes allow every benign task and therefore have a false-positive rate of zero in this diagnostic suite. OAuth-only and Attest-on-connect block none of the six execution-misuse families. Stateful LP blocks five families but misses streaming semantic mutation because it lacks fresh workload and invocation-state binding. Capability-only blocks four families but misses tool-chain step injection and streaming semantic mutation. Full \system{} blocks every evaluated scenario in every repetition and has a false-negative rate of zero. These scenarios broaden the evaluation context without changing the paper's central claim about post-authorization workload trust.

\begin{table}[t]
\centering
\caption{Agent tool-use results over ten repetitions. Latency is the request-level pooled p95 over normal expected-allow requests that were allowed.}
\label{tab:agent-main}
\small
\setlength{\tabcolsep}{3.0pt}
\begin{tabular}{lrrr}
\toprule
Mode & Benign success & Attack block & Normal p95 \\
 & & & (ms) \\
\midrule
OAuth-only & 1.000 & 0.0000 & 12.20 \\
Attest-on-connect & 1.000 & 0.0000 & 17.30 \\
Stateful LP & 1.000 & 0.8333 & 11.06 \\
Capability-only & 1.000 & 0.6667 & 11.57 \\
\textbf{ACLE-MCP} & \textbf{1.000} & \textbf{1.0000} & \textbf{15.34} \\
\bottomrule
\end{tabular}
\end{table}

\begin{figure}[t]
\centering
\includegraphics[width=\columnwidth]{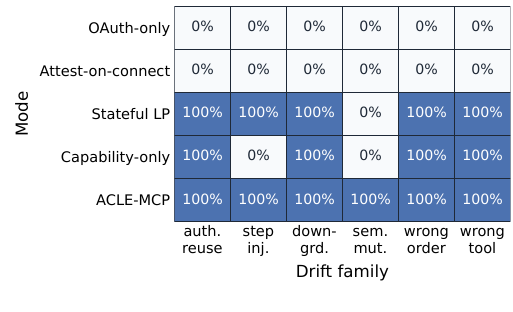}
\caption{Attack-blocking rates in the agent tool-use extension. Full ACLE-MCP blocks all six execution-misuse families, whereas weaker modes cover only subsets.}
\label{fig:agent-families}
\end{figure}

In the current local simulation, OAuth-only and the complete mechanism each contribute 40 normal allowed request samples. The complete mechanism increases request-level pooled p95 latency from 12.20~ms to 15.34~ms, an increase of 3.14~ms or 25.7\% (1.26$\times$). This result reflects the current warm-cache simulated Verifier path; it does not represent cold-cache operation, real JWKS retrieval, or hardware-attestation cost. Four benign task families are insufficient to support a broad claim that general agent capability is unaffected. The result shows only that the observed security gains are not achieved through indiscriminate denial.

\subsection{RQ2: Ablation Study and Mechanism Attribution}
We repeat the complete \system{} configuration and its component-removal variants over the same 24 scenarios, 10 random seeds, and single-concurrency setting. Each configuration contains 190 attack samples expected to be blocked and 30 benign samples. Table~\ref{tab:ablation} reports the blocking rate over attacks expected to be denied, rather than mixing attack-labeled control cases that are expected to be allowed into the denominator. Under this definition, the complete design achieves a 100\% blocking rate.

Removing sender proof admits token replay and cross-client lease transfer. Removing freshness checks reopens TOCTOU and stale appraisal-result reuse. Removing downstream constraints reopens undeclared proxying and SSRF access to metadata services. Removing receipt validation admits missing, forged, and digest-mismatched receipts. The configuration without Attested Step-Up also disables workload-binding checks and should therefore be interpreted as a joint ablation of these two mechanisms rather than as removal of the Verifier alone. All security ablations preserve 100\% benign-task success, indicating that the degradation arises from missing checks rather than indiscriminate blocking of normal requests.

\begin{table*}[t]
\centering
\caption{Component ablations over ten repetitions. Blocking rate uses the 190 attacks expected to be denied as the denominator; p95 includes only normal allowed requests.}
\label{tab:ablation}
\small
\setlength{\tabcolsep}{3.0pt}
\begin{tabularx}{\textwidth}{@{}lXrrrrrX@{}}
\toprule
Configuration & Removed component & Block rate & FN & Benign success & Normal p95 & Verifier calls & Reopened attack families \\
 & & (\%) & & (\%) & (ms) & & \\
\midrule
\texttt{C\_full} & None (complete design) & 100.00 & 0 & 100.00 & 21.84 & 27 & None \\
\texttt{C\_no\_sender} & Sender proof/binding & 89.47 & 20 & 100.00 & 21.56 & 30 & Token replay; cross-client lease transfer \\
\texttt{C\_no\_freshness} & Freshness checks & 89.47 & 20 & 100.00 & 37.32 & 29 & TOCTOU; stale appraisal reuse \\
\texttt{C\_no\_downstream} & Downstream constraints & 84.21 & 30 & 100.00 & 21.19 & 35 & Undeclared proxy; SSRF metadata access \\
\texttt{C\_no\_receipt} & Receipt requirement/validation & 78.95 & 40 & 100.00 & 20.78 & 25 & Missing, forged, or digest-mismatched receipts \\
\texttt{C\_no\_attest\_stepup} & Attested Step-Up and workload binding & 84.21 & 30 & 100.00 & 20.84 & 0 & Compromised, substituted, or mismatched workloads \\
\texttt{C\_cache\_off} & AR cache & 100.00 & 0 & 100.00 & 23.45 & 210 & None; only higher Verifier load and normal-path cost \\
\bottomrule
\end{tabularx}
\end{table*}

\begin{table}[ht]
\centering
\caption{Real-component and backend-substitution validation. ``Pass'' denotes successful integration checks, not production-grade performance.}
\label{tab:integration}
\small
\setlength{\tabcolsep}{3.0pt}
\begin{tabularx}{\columnwidth}{@{}>{\raggedright\arraybackslash}X>{\raggedright\arraybackslash}Xc@{}}
\toprule
Validation target & Exercised property & Result \\
\midrule
Keycloak/OIDC & Issuer, JWKS, client, and audience checks & Pass \\
MCP Python SDK & Protected tool invocation after gate admission & Pass \\
vTPM quote path & Quote material and nonce-bound verification & Pass \\
Full control chain & Verifier, lease-issuer, and gate interfaces & Pass \\
\bottomrule
\end{tabularx}
\end{table}

The complete design and the configuration with the appraisal-result cache disabled have identical security outcomes. Disabling the cache increases Verifier calls from 27 to 210 and raises normal pooled p95 latency from 21.84~ms to 23.45~ms, an increase of approximately 7.4\%. The AR cache is therefore a performance optimization rather than a security requirement. The main ablation uses single concurrency; although disabling single-flight leaves security unchanged, this setting does not actually exercise concurrent-request coalescing, so we do not quantify the performance benefit of single-flight.

\begin{figure}[t]
    \centering
    \includegraphics[width=\columnwidth]{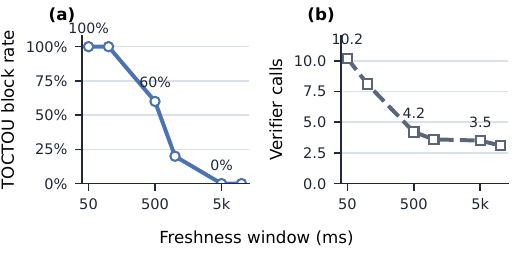}
    \caption{Security-performance trade-off of appraisal freshness: (a) TOCTOU blocking rate under different freshness windows; and (b) the corresponding number of Verifier calls. Wider reuse windows reduce verification pressure but enlarge the exposure interval between appraisal and execution.}
    \label{fig:freshness}
\end{figure}

\paragraph{Freshness sensitivity.}
Freshness is not merely a performance parameter. We hold the complete enforcement mode fixed and sweep the maximum acceptable age of an appraisal result. Figure~\ref{fig:freshness} shows that a 50~ms window blocks all modeled TOCTOU cases, with an average of 10.2 Verifier calls and a cache-hit ratio of 0.5143. At 500~ms, the TOCTOU blocking rate falls to 0.60; at 5000~ms, it falls to zero. This result shows why connect-time attestation is insufficient for later high-risk invocations. Such operations require short freshness windows, active invalidation when the workload changes, or single-use leases. Wider reuse windows are appropriate only when the operation policy explicitly accepts the corresponding exposure.

\begin{figure}[!t]
\centering
\includegraphics[width=\columnwidth]{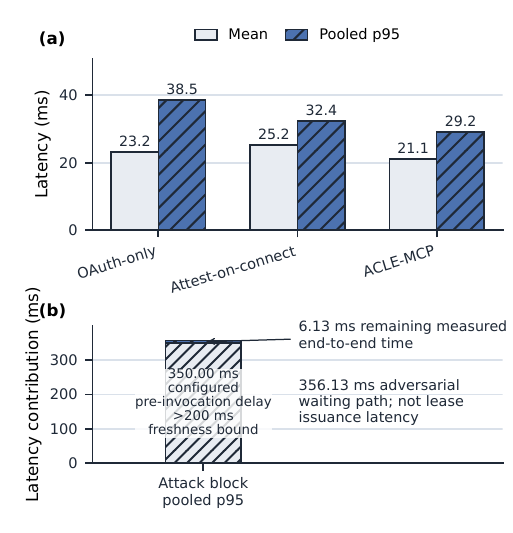}
\caption{Mechanism-level latency. (a) Mean and pooled p95 for normal requests. (b) The 356.13~ms TOCTOU block comprises a configured 350~ms delay and 6.13~ms residual time, not lease or gate overhead.}
\label{fig:path-latency}
\end{figure}

\subsection{RQ3: Real-Component Integration and Runtime Cost}
Full \system{} has a request-level pooled p95 of 15.34~ms for normal allowed requests in the agent suite, compared with 12.20~ms for OAuth-only, a relative increase of 25.7\%. The mechanism-level experiment further separates normal and blocking paths in Figure~\ref{fig:path-latency}. Normal allowed requests have a pooled mean of 23.17~ms and p95 of 38.41~ms. The 356.13~ms blocking-path p95 is dominated by a configured 350~ms pre-invocation delay that makes the appraisal exceed the 200~ms freshness threshold; it is not lease-issuance or gate overhead. The agent and mechanism-level experiments use different scenario mixtures and should not be numerically combined.

The Keycloak/MCP path validates a real Keycloak token, enforces the intended resource audience at the \gate{}, and invokes an MCP Python SDK tool. It does not yet include production remote attestation, DPoP-bound issuance, or a complete hardware trust chain. Table~\ref{tab:integration} summarizes the exercised components. The optional vTPM path replaces simulated evidence signatures with nonce-bound quote generation and verification while preserving the appraisal-result, lease-issuance, and gate interfaces.

\section{Discussion}
\label{sec:Discussion}
\system{} is a reference-monitor architecture for the \trustgap{}, not a complete trust framework for agentic systems. Its key assumption is a non-bypassable provider-side \gate{}: protected requests must reach the application workload only after admission. If an attacker controls the gate or all routing paths, enforcement fails. The prototype also excludes prompt injection, incorrect Host-side planning, and compromise of the Verifier or lease issuer.

The evaluation is limited to four benign and six author-constructed misuse families. Sender proof and workload appraisal are simulated in the main harness, while the vTPM path verifies quotes without binding every workload claim. Receipts cannot undo external side effects, and downstream guarantees cover only declared or mediated dependencies. These results therefore demonstrate prototype-level feasibility rather than production readiness.

\section{Conclusion} \label{sec:Conclusion} We studied the \trustgap{} in remote LLM tool use: OAuth can authorize a client to access an MCP service without ensuring that a later invocation is executed by the expected and freshly appraised provider-side workload. \system{} closes this gap by issuing a short-lived, sender-constrained capability lease that binds workload appraisal to the invocation boundary and is consumed immediately before protected tool logic begins. Controlled attacks, component ablations, and the agent tool-use extension show that authorization-only, connect-time attestation, Stateful policy, and Capability-only designs leave distinct attack surfaces, whereas the complete design preserves the evaluated benign tasks and blocks all evaluated attack families. The conclusion is intentionally narrow: for high-risk remote tool calls, workload trust must be checked at the point where delegated authority is actually consumed.

\bibliography{aclemcp}

% If so, uncomment the following line and adjust the path.

\end{document}